\documentclass[12pt]{article}

\usepackage{amsmath}
\usepackage{amssymb}
\usepackage{mathrsfs}
\usepackage[dvips]{graphicx}

\newcommand{\beq}{\begin{equation}}
\newcommand{\bea}{\begin{eqnarray}}
\newcommand{\eeq}{\end{equation}}
\newcommand{\eea}{\end{eqnarray}}

\newcommand{\bp}{{\mathbf p}}
\newcommand{\bq}{{\mathbf q}}
\newcommand{\bl}{{\mathbf l}}
\newcommand{\br}{{\mathbf r}}

\newcommand{\bx}{{\mathbf x}}

\newcommand{\bj}{{\mathbf j}}

\graphicspath{{fig/}}

\begin{document}

\title{\bf Strong versus weak wave-turbulence in relativistic field theory}

\author{
J{\"u}rgen Berges$^1$, D\'enes Sexty$^2$\\[0.5cm]
$^1$Institute for Nuclear Physics\\
Darmstadt University of Technology\\
Schlossgartenstr. 9, 64289 Darmstadt, Germany\\
$^2$Institute for Theoretical Physics\\
University of Heidelberg\\
Philosophenweg 16, 69120 Heidelberg, Germany}

\date{}
\begin{titlepage}
\maketitle
\def\thepage{}          % No page number on title page

\begin{abstract}
Nonthermal scaling phenomena can exhibit a characteristic dependence on the dimensionality $d$ of space. For $d=3$ and $4$ we simulate a relativistic scalar field theory on a lattice and compute turbulent scaling exponents. We recover Kolmogorov or weak wave-turbulence in the perturbative high-momentum regime, where it exhibits the scaling exponent $\kappa_{\mathrm w} = d - 3/2$. In the nonperturbative infrared regime, we find a different scaling exponent $\kappa_{\mathrm s} = 4\, (5)$ for $d=3\, (4)$, which is in agreement with the recently predicted anomalously large values $\kappa_{\mathrm s} = d + 1$ of strong turbulence. We show how the latter can be seen to characterize stationary transport of a conserved effective particle number.  
\end{abstract}

\end{titlepage}

\renewcommand{\thepage}{\arabic{page}}

\section{Introduction}

Turbulent scaling phenomena appear for very different physical systems and length scales, ranging from early universe reheating dynamics~\cite{Micha:2002ey,Berges:2008wm} to physics of supernova explosions~\cite{supernovae} or laboratory experiments with ultra-cold quantum gases~\cite{Scheppach:2009wu}. While many aspects of turbulence have long reached textbook level~\cite{Kolmogorov}, rather little is known about turbulent behavior in nonperturbative regimes of relativistic quantum field theories. Here the strong interest is to a large extent also driven by related questions concerning relativistic heavy-ion collisions~\cite{HIC}. 

It has recently been demonstrated that a new class of turbulent scaling phenomena exist in the nonperturbative regime of relativistic scalar field theories for sufficiently high occupation numbers per mode~\cite{Berges:2008wm}. The nonthermal scaling solutions were shown to exhibit a strong enhancement $\sim |{\bf p}|^{-4}$ in the infrared for three spatial dimensions, while at high momenta a well-known scaling regime of weak wave-turbulence $\sim |{\bf p}|^{-3/2}$ is observed. In Ref.~\cite{Berges:2008sr} a characteristic dependence of the nonperturbative infrared solutions on the dimensionality of space was predicted. Turbulent scaling phenomena are insensitive to the details of the underlying microscopic theory. In particular, the  infrared scaling solutions are expected to belong to universality classes. These may only depend on few general properties such as space dimension, symmetry, field content and conserved charges.
  
In this work we compute turbulent scaling behavior of a self-interacting $N$-component scalar field theory. This is done using classical-statistical simulations on a lattice, which are expected to be an accurate description also for the corresponding quantum theory for the considered high occupation numbers per mode~\cite{Berges:2008wm}. In order to further classify the universality class of nonperturbative scaling solutions, we perform simulations in three and four spatial dimensions. In addition, we simulate for $N=4$ and $N=10$ field components. We find striking agreement of the numerical results with the previously predicted analytic dependence on the dimensionality of space based on resummed large-$N$ techniques~\cite{Berges:2008sr}. In particular, we find no indication for a dependence on $N$ and the results are consistent with a vanishing anomalous dimension as well as a dynamic scaling exponent $z=1$ for the relativistic theory. 
We show that the phenomenon of strong turbulence at low momenta may be associated to stationary transport of a conserved effective particle number. 

The paper is organized as follows. We start with the definition of suitable correlation functions, which can be used to discuss weak as well as strong turbulence from a common framework in Sec.~\ref{sec:analytic}. After a brief review of perturbative wave-turbulence using kinetic theory, we analytically discuss strong turbulence as the stationary transport of a conserved effective particle number. In Sec.~\ref{sec:lattice} we discuss the numerical results from lattice simulations. 

\section{Stong versus weak stationary turbulence}
\label{sec:analytic}

We describe turbulence in a relativistic self-interacting scalar field theory as the stationary transport of a conserved quantity. Typically this is done perturbatively using kinetic theory. To obtain a description that is valid also beyond perturbation theory, we 
first define correlations functions which suitably describe the physics of turbulence. These are evaluated perturbatively using kinetic theory in Sec.~\ref{sec:pert} and then nonperturbatively in Secs.~\ref{sec:nonpert} and \ref{sec:lattice}.

For a real scalar field theory with Heisenberg field operator $\Phi(x)$, where $x = (x^0,\bx)$ denotes the time $x^0$ and $d$-dimensional space variable $\bx$, we consider the expectation value of the anti-commutator $F(x,y)$ and commutator $\rho(x,y)$ of two fields,
\begin{equation}
F(x,y) = \frac{1}{2}\langle \{\Phi(x),\Phi(y)\} \rangle \quad , \quad 
\rho(x,y) = i \langle [\Phi(x),\Phi(y)] \rangle \, , 
\label{eq:Frhodef}
\end{equation}
respectively. Here the real spectral function $\rho(x,y)$ is related to the retarded or advanced propagators by
\begin{equation}
G^R(x,y) \, = \, G^A(y,x) \, = \, \Theta (x^0 - y^0)\, \rho(x,y) \, ,
\label{eq:retadv}
\end{equation}
with the step function $\Theta (x^0 - y^0)=1$ for $x^0 > y^0$ and zero otherwise. Because of anti-symmetry, $\rho(x,y) = - \rho (y,x)$, one has $\rho(x,x) = 0$. The real statistical two-point function is symmetric, $F(x,y) = F(y,x)$. It is convenient to introduce Wigner coordinates 
\begin{equation}
X^\mu = \frac{x^\mu + y^\mu}{2} \quad , \quad s^\mu = x^\mu - y^\mu
\end{equation}
and Fourier transform with respect to the relative coordinates,
\begin{eqnarray}
F_p(X) = \int {\mathrm d}^{d+1} s\, e^{-ip_\mu s^\mu} F(X+s/2, X-s/2) \, ,
\\
\tilde{\rho}_p(X) = -i \int {\mathrm d}^{d+1} s\, e^{-ip_\mu s^\mu} \rho(X+s/2, X-s/2) \, 
\end{eqnarray}
with $p = (p^0,\bp)$. The conventional factor of $-i$ in the transform for the spectral function makes the latter real in Fourier space and we use a tilde to denote this. According to (\ref{eq:retadv}), retarded and advanced propagators in Fourier space are then related to the spectral function by
\begin{equation}
i\tilde{\rho}_p(X) \, = \, G^R_p(X) - G^A_p(X) \, . 
\end{equation}
For the real scalar field theory one finds from the definitions (\ref{eq:Frhodef}),
\begin{equation}
F_{-p}(X) \, = \, F_p(X) \quad , \quad \tilde{\rho}_{-p}(X) \, = \, - \tilde{\rho}_p(X) \, .
\label{eq:symFrho}
\end{equation}
Without loss of generality we write
\begin{equation}
F_p(X) \, = \, \left( n_p(X) + \frac{1}{2} \right) \tilde{\rho}_p(X) \, ,
\label{eq:defn}
\end{equation}
which defines the function $n_p(X)$ for any given $F_p(X)$ and $\tilde{\rho}_p(X)$. We emphasize that without additional assumptions (\ref{eq:defn}) does not represent a fluctuation-dissipation relation, which holds only if $n_p(X)$ is replaced by a thermal distribution function. We will not assume this in the following and keep $n_p(X)$ general at this stage. In particular,
(\ref{eq:symFrho}) then implies the identity
\begin{equation}
n_{-p}(X) \, = \, - \left( n_p(X) + 1 \right) \, .
\end{equation}

Since we are interested in stationary behavior, we may consider the dynamics employing a gradient expansion to lowest order in the number of derivatives with respect to the center coordinates $X^\mu$ and powers of the relative coordinates $s^\mu$. This is a standard procedure for the derivation of kinetic equations from field theory, and for the spectral function to lowest order one has~\cite{Berges:2005md,Blaizot:2001nr} 
\begin{equation}
2 p^\mu \frac{\partial}{\partial X^\mu} \tilde{\rho}_p(X) = 0 \, .
\label{eq:rhograd}
\end{equation}
For spatially homogeneous ensembles (\ref{eq:rhograd}) implies a constant $\tilde{\rho}_p$ that does not depend on time. In contrast, the statistical function $F_p(t)$ to this order can depend on time $t \equiv X^0$ and using (\ref{eq:defn}) we can write
\begin{equation}
\int_0^\infty \frac{{\mathrm d} p^0}{2\pi}\, 2 p^0 \frac{\partial}{\partial t} F_p(t) \, = \, 
\int_0^\infty \frac{{\mathrm d} p^0}{2\pi}\, 2  p^0 \tilde{\rho}_p\, \frac{\partial n_p(t)}{\partial t} \, = \, C[n](t;\bp) \, .
\label{eq:genev} 
\end{equation}
Here $C[n]$ denotes the 'gain' and 'loss' terms, which describe the effects of interactions to lowest order in the gradient expansion. We will determine $C[n]$ using perturbation theory in Sec.~\ref{sec:pert}, and in Sec.~\ref{sec:nonpert} it will be obtained from a resummed $1/N$ expansion to next-to-leading order (NLO). 

\subsection{Weak wave-turbulence}
\label{sec:pert}

In this section we review some relevant aspects of perturbative Kolmogorov or weak wave-turbulence~\cite{Turbulence}, which will be used below for comparison with the nonperturbative regime of strong turbulence. The free spectral function is
\begin{equation}
\tilde{\rho}^0_p \, = \, 2 \pi\, {\mathrm{sgn}}(p^0) \, \delta\!\left((p^0)^2-\omega_\bp^2\right)
\label{eq:freerho}
\end{equation}
for a relativistic scalar field theory with particle energy $\omega_\bp$. Plugging this into (\ref{eq:genev}) gives
\begin{equation}
\frac{\partial n_\bp(t)}{\partial t} 
\,=\, C[n] (t;\bp) \, , 
\label{eq:kinetic}
\end{equation}
where
\begin{equation}
n_\bp (t) \equiv \int_0^\infty \frac{{\mathrm d} p^0}{2\pi}\, 2  p^0 \tilde{\rho}^0_p\, n_p(t) \, .
\end{equation}
For a compact notation we will frequently suppress the $t$-dependence and write $n_\bp$.
In kinetic theory, when two particles scatter into two particles, the collision integral on the RHS of (\ref{eq:kinetic}) is of the form
\begin{equation}
C^{2\leftrightarrow 2}(\bp) =  \int {\mathrm{d}}\Omega^{2\leftrightarrow 2}(\bp,\bl,\bq,\br)
\left[(1+n_\bp) (1+n_\bl) n_\bq n_\br - n_\bp n_\bl (1+n_\bq) (1+n_\br)\right]
\label{eq:2to2}
\end{equation}
where the gain and loss terms in the integrand take into account that for bosons there is an enhancement of the rate if the final state is already occupied. The details of the model enter $\int {\mathrm{d}}\Omega^{2\leftrightarrow 2}(\bp,\bl,\bq,\br)$, which for an $O(N)$-symmetric scalar field with quartic $\lambda/(4!N)$-interaction reads~\cite{Berges:2004yj} 
\begin{eqnarray}
\int {\mathrm{d}}\Omega^{2\leftrightarrow 2}(\bp,\bl,\bq,\br) & = & \lambda^2\, \frac{N+2}{18 N^2} \int_{\bl \bq \br} \!\! (2\pi)^{d+1}\, \delta^{(d)}(\bp+\bl-\bq-\br) 
\nonumber\\
&\times& \delta(\omega_\bp + \omega_\bl - \omega_\bq - \omega_\br)\, \frac{1}{2\omega_\bp 2\omega_\bl 2\omega_\bq 2\omega_\br}
\label{eq:measurepert}
\end{eqnarray}
with the notation $\int_\bp \equiv \int {\mathrm{d}}^d p/(2\pi)^d$.
Using the approximation (\ref{eq:2to2}) with (\ref{eq:measurepert}) for the collision integral on the RHS of (\ref{eq:kinetic}), one obtains the well-known Boltzmann equation for a gas of relativistic particles. Clearly, this approximation cannot be used if the occupation numbers per mode become large and, parametrically, for a weak coupling $\lambda$ a necessary condition is $n_\bp \ll 1/\lambda$~\cite{Berges:2004yj}. In Sec.~\ref{sec:nonpert} we will discuss suitable approximations that are valid also for nonperturbatively large occupation numbers. 

Turbulence is expected for not too small occupation numbers per mode such that quantum corrections can be neglected. For a regime $1 \ll n_\bp \ll 1/\lambda$ one may still use the above Boltzmann equation, which becomes 
\begin{equation}
\frac{\partial n_\bp(t)}{\partial t} \, \simeq \, \int {\mathrm{d}}\Omega^{2\leftrightarrow 2}(\bp,\bl,\bq,\br) \left[ 
(n_\bp + n_\bl) n_\bq n_\br - n_\bp n_\bl (n_\bq + n_\br)\right] \, .
\label{eq:2to2cl}
\end{equation}
For the considered real scalar field theory the energy density $\epsilon$ is conserved. Since we restrict our discussion in this section to number conserving $2 \leftrightarrow 2$ scatterings also the total particle number $n_{\mathrm{tot}}$ is conserved, and
\begin{equation}
\epsilon \, = \, \int_\bp \omega_\bp n_\bp \quad , \quad n_{\mathrm{tot}} \, = \, \int_\bp n_\bp \, .
\label{eq:epsntot}
\end{equation}
The fact that they are conserved and interactions 
are sufficiently local in momentum space\footnote{Strongly non-local contributions 
are suppressed by phase space and energy-momentum conservation. The classical theory requires, of course, an ultraviolet cutoff to regularize the Rayleigh-Jeans divergence.}
may be described by a continuity equation in Fourier space, such as 
\begin{equation}
\frac{\partial}{\partial t}\left( \omega_\bp n_\bp \right) + \nabla_\bp \cdot \bj_\bp \, = \, 0
\label{eq:continuity}
\end{equation}
for energy conservation~\cite{Turbulence}. Similarly, particle number conservation is described by formally replacing $\omega_\bp \to 1$ in the above equation and corresponding 
substitution of the flux density. For isotropic ensembles we consider the energy flux $A(k)$ through a momentum sphere of radius $k$. Then only the radial component of the flux density $\bj_\bp$ is nonvanishing and
\begin{equation}
\int_\bp^k \nabla_\bp \cdot \bj_\bp \, = \, \int_{\partial k} \bj_\bp \cdot {\mathrm d}{\mathbf{A}}_\bp \, \equiv \, (2\pi)^d A(k)\, .
\end{equation}
Since $\omega_\bp$ is constant in time, we can thus write with the help of (\ref{eq:continuity})  
\begin{equation}
A(k) \, = \, - \frac{1}{2^d \pi^{d/2} \Gamma(d/2+1)}\, \int^k \! {\mathrm d}p \, |\bp|^{d-1} \omega_\bp \frac{\partial n_\bp(t)}{\partial t} \, .
\label{eq:flux}
\end{equation}
For stationary turbulence the flux $A(k)$ is scale independent, i.e.~the respective integral does not depend on the integration limit $k$. We consider scaling solutions 
\begin{equation}
n_{\bp} \, = \, s^{\kappa_{\mathrm w}} n_{s\bp} \quad , \quad 
\omega_{\bp} \, = \, s^{-1}\, \omega_{s\bp} \, ,
\label{eq:scalingprop}
\end{equation}
with occupation number exponent $\kappa_{\mathrm w}$ and assuming a linear dispersion relation.
Since the physics is scale invariant, we can choose $s = 1/|\bp|$ such that $n_\bp = |\bp|^{-\kappa_{\mathrm w}}\, n_1$ and $\omega_\bp = |\bp|\, \omega_1$.
Using the scaling properties (\ref{eq:scalingprop}) one obtains from (\ref{eq:measurepert})
\begin{equation}
\int \mathrm{d} \Omega^{2\leftrightarrow 2}(\bp,\bl,\bq,\br) \, = \, s^{-\mu_4}\, \int \mathrm{d} \Omega^{2\leftrightarrow 2}(s\bp,s\bl,s\bq,s\br) \, . 
\end{equation}
Here the scaling exponent $\mu_4$ for the theory with quartic self-interaction is given by 
\begin{equation}
\mu_4 \, = \, (3d-4)-(d+1) \, = \, 2d - 5 \, ,
\label{eq:Del4}
\end{equation}
where the first term in brackets comes from the scaling of the measure and the second from energy-momentum conservation for two-to-two scattering described by (\ref{eq:2to2}). Apart from the quartic self-interaction, it will be relevant to consider also scattering in the presence of a non-vanishing field expectation value such that an effective $3$-vertex appears~\cite{Berges:2004yj}. In this case, one obtains along these
lines
\begin{equation}
\mu_3 \, = \, (2d-3)-(d+1) \, = \, d - 4 \, .
\end{equation}
To keep the discussion more general, we write for the scaling properties of the flux for a given $m$-vertex
\begin{eqnarray}
A(k) & = & - \frac{1}{2^d \pi^{d/2} \Gamma(d/2+1)}\, \int^k \! {\mathrm d}p\, |\bp|^{d-1+1-\kappa_{\mathrm w}(m-1)+\mu_m} \, \omega_1  
\frac{\partial n_1}{\partial t} \, .
\label{eq:intA}
\end{eqnarray}
For $m=4$ this can be directly verified to agree to the two-to-two scattering case with a $4$-vertex using (\ref{eq:2to2cl}) and (\ref{eq:Del4}).
If the exponent in the integrand of (\ref{eq:intA}) is nonvanishing, the integral gives
\begin{equation}
A(k) \, \sim \,  \frac{k^{d+1-\kappa_{\mathrm w}(m-1)+\mu_m}}{d+1-\kappa_{\mathrm w}(m-1) + \mu_m}\, \omega_1 \frac{ \partial n_1}{\partial t} \, .
\end{equation} 
Scale invariance up to logarithmic corrections is, therefore, obtained for
\begin{equation}
d+1-\kappa_{\mathrm w}(m-1)+\mu_m \, = \, 0 \, .
\end{equation}
This yields for the energy cascade
\begin{equation}
\kappa_{\mathrm w} \, \stackrel{m=4}{=} \, d - \frac{4}{3} \quad , \quad  
\kappa_{\mathrm w} \, \stackrel{m=3}{=} \, d - \frac{3}{2} \, .
\label{eq:ekappaweak}
\end{equation}
One observes that stationary turbulence requires the existence of the limit
\begin{equation}
\lim_{d+1-\kappa_{\mathrm w}(m-1)+\mu_m \to 0} \,\, \frac{\partial n_1/\partial t}{d+1-\kappa_{\mathrm w}(m-1)+\mu_m} \, = \, {\mathrm{const}} \, \neq \, 0 \, ,
\end{equation}
such that the collision integral must have a corresponding zero of
first degree. Similarly, starting from the continuity equation for
particle number one can study stationary turbulence associated to
particle number conservation. For instance, for two-to-two scattering
this leads to $\kappa_{\mathrm w} = d - 5/3$, and $ \kappa_{\mathrm w}
= d - 2 $ for the case of interaction through a 3-vertex.

\subsection{Strong turbulence}
\label{sec:nonpert}

The above perturbative description of stationary turbulence becomes invalid at low momenta $|\bp|$, since the occupation numbers $n_\bp \sim |\bp|^{-\kappa_{\mathrm{w}}}$ can grow nonperturbatively large in the infrared. This concerns positive values of the scaling exponent $\kappa_{\mathrm{w}}$ given by (\ref{eq:ekappaweak}), which is the case for the dimensions $d=3$ and $4$ to be considered below. To understand where the picture of weak wave-turbulence breaks down and to compute the properties of the infrared regime, we have to consider nonperturbative approximations. In this section we consider the expansion of the two-particle irreducible (2PI) effective action in the number of field components to NLO to get analytical insight~\cite{Berges:2001fi}. We extend the discussions of Refs.~\cite{Berges:2008wm,Berges:2008sr} by showing that a conserved effective particle number characterizes strongly modified scaling properties in the nonperturbative low-momentum regime.

At NLO in the 2PI $1/N$ expansion the evolution equation (\ref{eq:genev}) for
\begin{equation}
n_{{\mathrm{eff}}}(t;\bp) \, \equiv \, \int_0^\infty \frac{{\mathrm d} p^0}{2\pi}\, 2  p^0 \tilde{\rho}_p\, n_p(t) 
\label{eq:neff}
\end{equation} 
reads
\begin{equation}
\frac{\partial n_{{\mathrm{eff}}}(t;\bp)}{\partial t} \, = \, C^{{\mathrm{NLO}}}[n](t;\bp) \, .
\end{equation}
Here the NLO contribution~\cite{Berges:2001fi} can be written as
\begin{eqnarray} 
C^{{\mathrm{NLO}}}(\bp) \!\!\! & = & \!\! \int\! {\mathrm{d}}\Omega^{2\leftrightarrow 2}(p,l,q,r)  \left[ 
(1+n_p) (1+n_l) n_q n_r - n_p n_l (1+n_q) (1+n_r) \right] 
\nonumber\\
&+& \!\! \int\! {\mathrm{d}}\Omega^{1\leftrightarrow 3}_{(a)}(p,l,q,r)  \left[ 
(1+n_p) (1+n_l) (1 + n_q) n_r - n_p n_l n_q (1+n_r) \right]
\nonumber\\
&+& \!\! \int \! {\mathrm{d}}\Omega^{1\leftrightarrow 3}_{(b)}(p,l,q,r)  \left[ 
(1+n_p) n_l n_q n_r - n_p (1 + n_l) (1+n_q) (1+n_r) \right]
\nonumber\\
&+& \!\! \int \! {\mathrm{d}}\Omega^{0\leftrightarrow 4}(p,l,q,r)  \left[ 
(1+n_p) (1+n_l) (1+n_q) (1+n_r) - n_p n_l n_q n_r \right]
\nonumber\\
\label{eq:CNLO}
\end{eqnarray} 
where we consider the case of a vanishing macroscopic field, i.e.\ $\langle \Phi \rangle = 0$, relevant for the infrared regime~\cite{Berges:2008wm}. Again we suppress in the notation the time dependence. Here 
\begin{eqnarray}
\int {\mathrm{d}}\Omega^{2\leftrightarrow 2}(p,l,q,r) & = & \frac{\lambda}{18 N} \int_0^\infty \frac{{\mathrm{d}}p^0{\mathrm{d}}l^0{\mathrm{d}}q^0{\mathrm{d}}r^0}{(2\pi)^{4-(d+1)}} \int_{\bl \bq \br} \delta^{(d+1)}(p+l-q-r)
\nonumber\\ 
&\times& \tilde{\rho}_p \tilde{\rho}_l \tilde{\rho}_q \tilde{\rho}_r \left[ \lambda_{\mathrm{eff}}(p+l) + \lambda_{\mathrm{eff}}(p-q) + \lambda_{\mathrm{eff}}(p-r) \right] \, ,
\nonumber\\
\int {\mathrm{d}}\Omega^{1\leftrightarrow 3}_{(a)}(p,l,q,r) & = & \frac{\lambda}{18 N} \int_0^\infty \frac{{\mathrm{d}}p^0{\mathrm{d}}l^0{\mathrm{d}}q^0{\mathrm{d}}r^0}{(2\pi)^{4-(d+1)}} \int_{\bl \bq \br} \,  \delta^{(d+1)}(p+l+q-r) 
\nonumber\\
&\times& \tilde{\rho}_p \tilde{\rho}_l \tilde{\rho}_q \tilde{\rho}_r \left[ \lambda_{\mathrm{eff}}(p+l) + \lambda_{\mathrm{eff}}(p+q) + \lambda_{\mathrm{eff}}(p-r) \right] \, ,
\nonumber\\
\int {\mathrm{d}}\Omega^{1\leftrightarrow 3}_{(b)}(p,l,q,r) & = & \frac{\lambda}{18 N} \int_0^\infty \frac{{\mathrm{d}}p^0{\mathrm{d}}l^0{\mathrm{d}}q^0{\mathrm{d}}r^0}{(2\pi)^{4-(d+1)}} \int_{\bl \bq \br} \,  \delta^{(d+1)}(p-l-q-r) 
\nonumber\\
&\times& \tilde{\rho}_p \tilde{\rho}_l \tilde{\rho}_q \tilde{\rho}_r \, \lambda_{\mathrm{eff}}(p-l) \, , 
\nonumber\\ 
\int {\mathrm{d}}\Omega^{0 \leftrightarrow 4}(p,l,q,r) & = & \frac{\lambda}{18 N} \int_0^\infty \frac{{\mathrm{d}}p^0{\mathrm{d}}l^0{\mathrm{d}}q^0{\mathrm{d}}r^0}{(2\pi)^{4-(d+1)}} \int_{\bl \bq \br} \,  \delta^{(d+1)}(p+l+q+r)
\nonumber\\
&\times& \tilde{\rho}_p \tilde{\rho}_l \tilde{\rho}_q \tilde{\rho}_r \, \lambda_{\mathrm{eff}}(p+l) \, .
\label{eq:omegaNLO}
\end{eqnarray}
In contrast to (\ref{eq:measurepert}), the above expressions still contain the integrations over frequencies and spectral functions. We emphasize that the latter are, in general, not of the free field form (\ref{eq:freerho}) in the nonperturbative regime. No quasi-particle assumptions has been employed and the only approximations are the $1/N$ expansion to NLO and the gradient expansion underlying (\ref{eq:genev}). 

The effective momentum-dependent 'coupling' $\lambda_{\mathrm{eff}}(p)$ appearing at NLO in the 2PI $1/N$ expansion is given by~\cite{Berges:2008wm,Berges:2008sr}
\begin{equation}
\lambda_{\mathrm{eff}}(p) \, = \, \frac{\lambda}{|1+\Pi^R(p)|^2} \, ,
\label{eq:lambdaeff}
\end{equation}
which involves the squared absolute value $|1+\Pi^R(p)|^2 = [1+\Pi^R(p)][1+\Pi^A(p)]$ of the retarded or advanced self-energy 
\begin{equation}
\Pi^{R,A}(p) \, = \, \frac{\lambda}{3}\int_q \left( n_{p-q} + \frac{1}{2} \right) \tilde{\rho}_{p-q}\,  G^{R,A}_q \, .
\label{eq:RAselfenergy}
\end{equation}
One observes that for sufficiently large $p$, for which $\Pi^{R,A}(p) \ll 1$, the effective coupling (\ref{eq:lambdaeff}) approaches $\lambda$.  In this case the '$2 \leftrightarrow 2$' contribution of the first line in (\ref{eq:CNLO}) is reminiscent of the two-to-two scattering process described by the perturbative expression presented in (\ref{eq:2to2}). The main difference is that the latter assumes a $\delta$-like spectral function such that all momenta are on shell. Therefore, in the perturbative expression (\ref{eq:2to2}) off-shell processes involving the decay of one into three particles or corresponding $3 \to 1$ annihilation processes or even $0 \leftrightarrow 4$ processes are absent. They can occur in principle at NLO in the 2PI $1/N$ expansion, which leads to the different terms contributing to the RHS of (\ref{eq:CNLO}). At sufficiently high momenta these off-shell contributions should be suppressed along with quantum-statistical corrections such that the spectral function approaches a $\delta$-like behavior. In this case we would recover stationary turbulence characterized by a weak scaling exponent $\kappa_{\mathrm{w}}$ described above. In contrast, in the infrared $\lambda_{\mathrm{eff}}(p)$ may have a nontrivial momentum dependence, which is discussed in the following.

Following similar lines as in Sec.~\ref{sec:pert}, we look for scaling solutions where $n_p \gg 1$ such that (\ref{eq:CNLO}) can be approximated by its classical-statistical limit
\begin{eqnarray}
C_{\mathrm{cl}}^{{\mathrm{NLO}}}(\bp) & = & \!\!\int {\mathrm{d}}\Omega^{2\leftrightarrow 2}(p,l,q,r)  \left[ 
(n_p + n_l) n_q n_r - n_p n_l (n_q + n_r) \right] 
\nonumber\\
&+& \!\! \int\! {\mathrm{d}}\Omega^{1\leftrightarrow 3}_{(a)}(p,l,q,r)  \left[ 
(n_p + n_l) n_q n_r - n_p n_l (n_q - n_r) \right]
\nonumber\\
&+& \!\! \int \! {\mathrm{d}}\Omega^{1\leftrightarrow 3}_{(b)}(p,l,q,r)  \left[ 
(-n_p + n_l) n_q n_r - n_p n_l (n_q + n_r) \right]
\nonumber\\
&+& \!\! \int \! {\mathrm{d}}\Omega^{0\leftrightarrow 4}(p,l,q,r)  \left[ 
(n_p + n_l) n_q n_r + n_p n_l (n_q + n_r) \right] \, .
\label{eq:CNLOcl}
\end{eqnarray} 
Again, only the first term above is reminiscent of the perturbative expression in (\ref{eq:2to2cl}). In principle, nonperturbative scaling phenomena may involve an anomalous scaling exponent for $\tilde{\rho}_p \equiv \tilde{\rho}(p^0,\bp)$. Using isotropy we write following Ref.~\cite{Berges:2008sr}
\begin{equation}
\tilde{\rho}(p^0,\bp) \, = \, s^{2-\eta}\, \tilde{\rho}(s^z p^0,s \bp) \, , 
\end{equation}
with a nonequilibrium 'anomalous dimension' $\eta$. A dynamical scaling exponent $z$ is taken into account since only spatial momenta are related by rotational symmetry and frequencies may scale differently because of the presence of (non-)thermal corrections. Scaling behavior of the statistical correlation function
\begin{equation}
F(p^0,\bp) \, = \, s^{2+\kappa_s}\, F(s^z p^0,s \bp)  
\end{equation}
then translates with (\ref{eq:defn}) for $n_p \gg 1$ into
\begin{equation}
n(p^0,\bp) \, = \, s^{\kappa_s + \eta}\, n(s^z p^0, s\bp) \, .
\end{equation}
With this one can determine the scaling behavior of $\lambda_{\mathrm{eff}}(p)$. From (\ref{eq:RAselfenergy}) follows
\begin{equation}
\Pi^{R,A}(p^0,\bp) \, = \, s^\Delta\, \Pi^{R,A}(s^z p^0, s\bp)
\end{equation}
with  
\begin{equation}
\Delta = 4 - d -z + \kappa_s - \eta .
\end{equation}
If $\Delta > 0$ one finds from (\ref{eq:lambdaeff}) the infrared scaling behavior
\begin{equation}
\lambda_{\mathrm{eff}}(p^0,\bp) \, = \, s^{-2\Delta}\, \lambda_{\mathrm{eff}}(s^z p^0, s\bp) \, . 
\end{equation}
For $\Delta \leq 0$ the effective coupling becomes trivial with $\lambda_{\mathrm{eff}}(p) \simeq \lambda$, on which we comment below. 
Using these scaling properties one obtains from (\ref{eq:omegaNLO})
\begin{equation}
\int {\mathrm{d}}\Omega^{2\leftrightarrow 2}(p,l,q,r) \, = \, s^{-2\kappa_s-z-2\eta} \int {\mathrm{d}}\Omega^{2\leftrightarrow 2}(s^z p^0,s^z l^0,s^z q^0,s^z r^0;s\bp,s\bl,s\bq,s\br)
\end{equation}
and the same scaling behavior for $\int {\mathrm{d}}\Omega^{1\leftrightarrow 3}_{(a)}(p,l,q,r)$, $\int {\mathrm{d}}\Omega^{1\leftrightarrow 3}_{(b)}(p,l,q,r)$ and $\int {\mathrm{d}}\Omega^{0\leftrightarrow 4}(p,l,q,r)$.

Similar to Sec.~\ref{sec:pert}, for any conserved quantity we can compute the flux through a momentum sphere $k$. Stationary turbulence solutions then require that the respective integral does not depend on $k$. Obviously, energy is conserved. The highly nontrivial question is whether a conserved effective particle number exists since there is no conserved charge associated to particle number in the real scalar field theory. Off-shell processes included in (\ref{eq:CNLOcl}), such as $1 \leftrightarrow 3$ processes, make this manifest. In the following, we analyze whether the effective particle number $n_{{\mathrm{eff}}}(t;\bp)$ given by $(\ref{eq:neff})$ represents a conserved quantity for the nonperturbative low-momentum regime. Similar to (\ref{eq:flux}), the flux for this effective particle number now reads 
\begin{equation}
A_{\mathrm{eff}}(k) \, = \, - \frac{1}{2^d \pi^{d/2} \Gamma(d/2+1)}\, \int^k \! {\mathrm d}p\, |\bp|^{d-1}\, \frac{\partial n_{{\mathrm{eff}}}(t;\bp)}{\partial t} \, .
\label{eq:Aeff}
\end{equation}
The momentum integral can be evaluated along the lines of Sec.~\ref{sec:pert} using the above scaling properties with
\begin{equation}
\frac{\partial n_{{\mathrm{eff}}}(t;\bp)}{\partial t} \, = \, |\bp|^{-\kappa_s+z-\eta} \,
\frac{\partial n_{{\mathrm{eff}}}(t;{\mathbf 1})}{\partial t} \, , 
\label{eq:scalneff}  
\end{equation}
such that 
\begin{equation}
A_{\mathrm{eff}}(k) \, \sim \,  \frac{k^{d-\kappa_s+z-\eta}}{d-\kappa_s+z-\eta}\, \frac{ \partial n_{{\mathrm{eff}}}(t;{\mathbf 1})}{\partial t} \, 
\end{equation} 
if the exponent in the integrand is nonvanishing. Scale invariance up to logarithmic corrections may, therefore, be obtained in the nonperturbative low-momentum regime for
\begin{equation}
\kappa_s \, = \, d + z - \eta \, .
\label{eq:kappas}
\end{equation}
This scaling solution is associated to a conserved $n_{{\mathrm{eff}}}(t;\bp)$ for sufficiently low momentum $\bp$.\footnote{Refs.~\cite{Berges:2008wm,Berges:2008sr} assume that the $1 \leftrightarrow 3$ and $0 \leftrightarrow 4$ contributions in (\ref{eq:CNLOcl}) vanish to obtain the solution (\ref{eq:kappas}). See also the discussion of this point in Ref.~\cite{Scheppach:2009wu}. We note that the momentum integral over $n_{{\mathrm{eff}}}(\bp)$ can be strongly infrared divergent for the discussed scaling solutions and requires an infrared cutoff.} Similarly, for the scaling solution associated to conserved energy one finds, taking into account an additional power of $p^0$, the exponent $\kappa_s = d+2z-\eta$ in accordance with Ref.~\cite{Berges:2008sr}.

The above discussion shows that in the presence of a conserved $n_{{\mathrm{eff}}}(t;\bp)$ there is a strongly modified infrared scaling behavior as compared to perturbative treatments. In particular, (\ref{eq:kappas}) predicts a characteristic dependence on the dimensionality of space $d$ and no dependence on the number $N$ of field components. 

\section{Lattice simulations}
\label{sec:lattice}

In this section we solve the evolution equations for our theory in the classical-statistical limit using simulations on a lattice. Varying the dimensionality of space $d$ and the number of field components $N$, we will then compare the numerical results with the analytical estimates (\ref{eq:kappas}) in the infrared and (\ref{eq:ekappaweak}) for high momenta.

The field equation of 
motion for the classical $N$-component scalar field theory reads with $ 1 \le a \le N $:
\begin{equation}
\ddot \varphi_a (t, \mathbf{x}) \, = \, \left( \bigtriangleup_\bx - m^2 \right) \varphi_a(t, \mathbf{x}) - \frac{\lambda}{6N}\, \sum_{b=1}^N \varphi_b(t, \mathbf{x})  \varphi_b(t, \mathbf{x}) \varphi_a (t, \mathbf{x}) \, .
\label{eq:classeom}
\end{equation}
For the numerical implementation of the above equation the leap-frog discretization is used on a cubic space-time lattice in three and four spatial dimensions. The initial conditions are chosen such that the system will evolve closely to non-thermal scaling solutions. To achieve this one can start with a nonequilibrium instability, such that low-momentum modes get highly populated~\cite{Berges:2008wm}. Such instabilities are, for example, the tachyonic instability or the parametric resonance instability, which also have cosmological relevance as models for reheating~\cite{Traschen:1990sw,Khlebnikov:1996mc,Berges:2002cz}. 
In this study we use initial conditions triggering parametric
resonance: the space average of the field has a nonzero initial value
$ \langle \varphi_1 (t=0) \rangle = \phi_0 $ while
$\langle \varphi_a (t=0) \rangle = 0 $ for $1 < a \le N $. The 
nonzero momentum modes are initialized with a small amplitude 
white noise\footnote{Its spectral composition is not important as long as the 
amplitude is small.} to provide a seed for unstable modes. 
The results are then averaged over different realizations of the 
initial noise distribution.

Our main observable, the momentum dependent particle number is defined by
\bea
n(t,\bp) = \frac{1}{N} \sum_{a=1}^N \sqrt { | \dot \varphi_a (t,\bp) | ^2  
| \varphi_a (t,\bp) | ^2 } \, ,
\eea
where $\varphi_a(t,\bp) $ is the spatial Fourier transform of the 
field in $d$ dimensions, i.e.~$\varphi_a(t,\bp)= 1/ \sqrt{V}  \int {\mathrm{d}}^d x  \varphi_a (t,\bx)  \exp( i \bp \bx )  $ with the spatial volume $V$. 

\begin{figure}
 \begin{center}
  \includegraphics[scale=0.9]{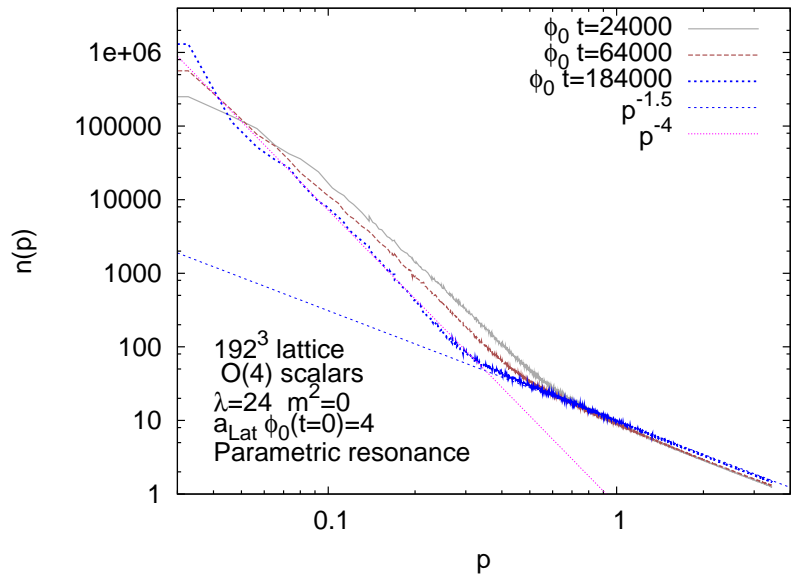}
  \caption[3d]{The particle number spectrum of the $d=3$ simulation for different
times in units of the initial field amplitude $\phi_0$.}
  \label{fig:3d}
 \end{center}
\end{figure}
Fig.~\ref{fig:3d} shows the particle number spectrum for a three 
dimensional simulation using a $192^3$ lattice with $\lambda=24,\ m^2=0$ and $N=4$.
The infrared modes exhibit a slow time evolution, whereas higher-momentum modes seem to settle much more quickly. For the final plotted time $\phi_0\, t=184000$ one observes two separate regions with clear power laws. For high momenta 
the scaling exponent $\kappa_w \simeq 1.5$ agrees well with the analytic prediction 
(\ref{eq:ekappaweak}) for Kolmogorov turbulence in three space dimensions as expected~\cite{Turbulence}. As the occupation number per mode grows towards lower momenta, the perturbative approximation breaks down for the description of the infrared modes. Accordingly, one observes a strongly modified power-law. Assuming $z=1$ for the relativistic theory and $\eta=0$, which is the case also at high momenta, the observed value $\kappa_s \simeq 4$ agrees well with the estimate $(\ref{eq:kappas})$ in accordance with the results of Ref.~\cite{Berges:2008wm}. 

\begin{figure}
 \begin{center}
  \includegraphics[scale=0.9]{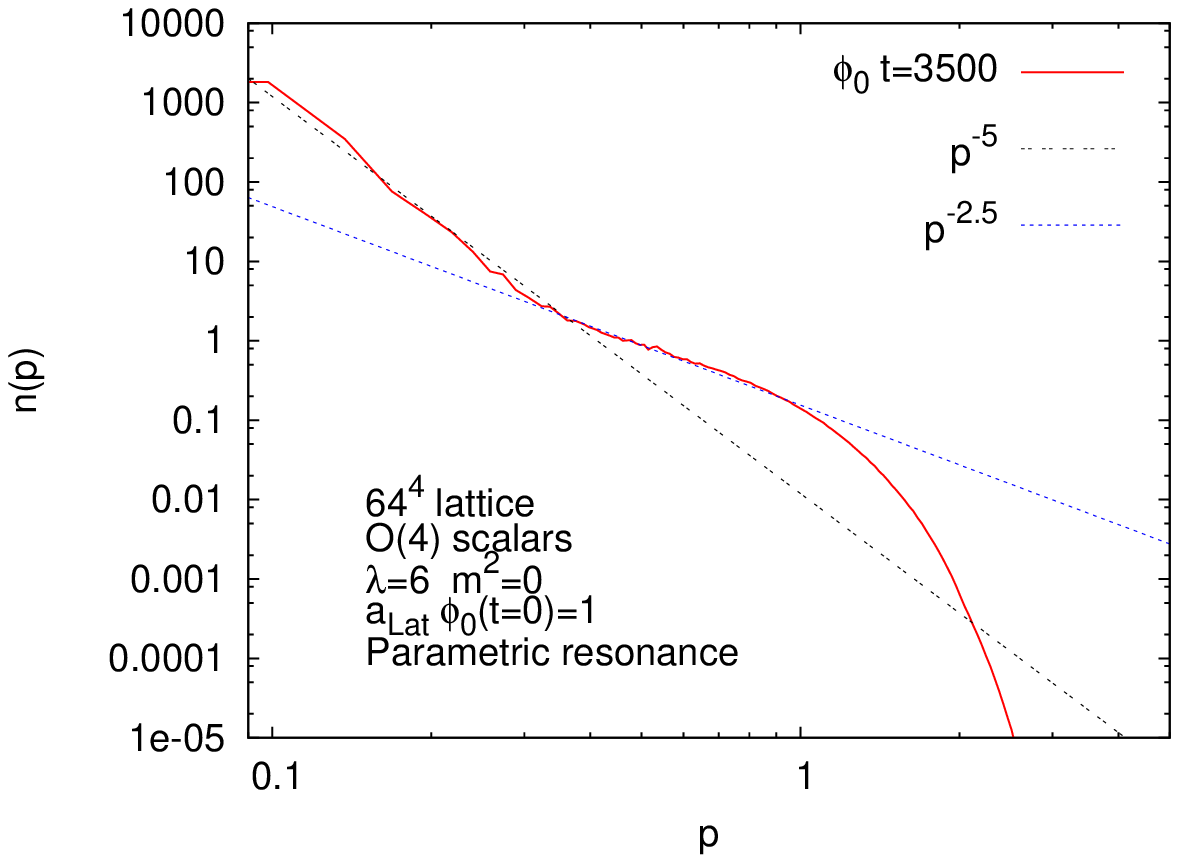}
  \includegraphics[scale=0.9]{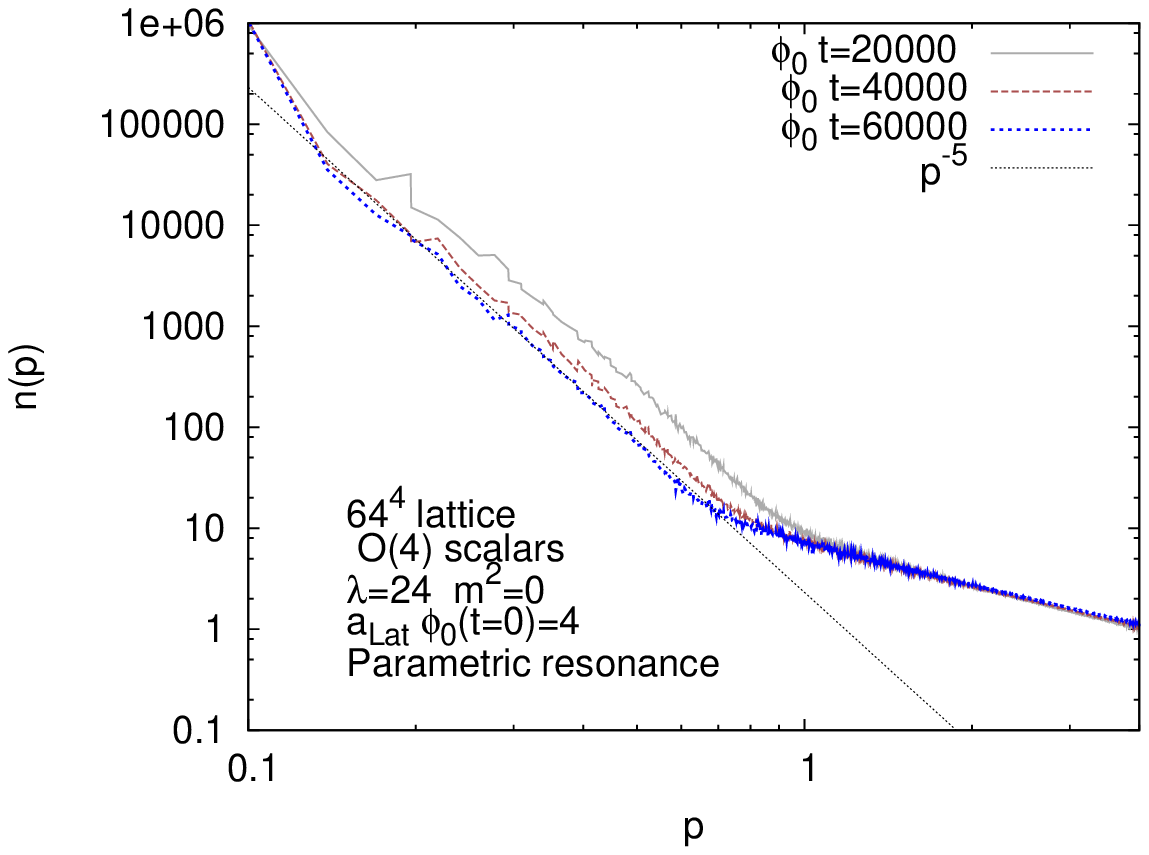}
  \caption[4d]{The particle number spectrum of the $d=4$ simulation
 for different times. The two graphs correspond to different initial 
conditions and different couplings as indicated. While the low-momentum scaling behavior is insensitive to these details, only the upper graph shows the expected Kolmogorov scaling at high momenta.}
  \label{fig:4d}
 \end{center}
\end{figure}
A crucial test for the interpretation of these results in terms of the analytic estimates of Secs.~\ref{sec:pert} and \ref{sec:nonpert} is their predicted characteristic momentum dependence. In Fig.~\ref{fig:4d} we show the particle 
number spectrum for simulations in $d=4$ for different initial conditions and coupling values. The upper graph indeed shows a low-momentum scaling exponent $\kappa_s \simeq 5$ as well as a high-momentum scaling exponent $\kappa_w \simeq 2.5$, which are in remarkable agreement with the predicted values $(\ref{eq:kappas})$ for $z=1$ and $\eta = 0$ as well as (\ref{eq:ekappaweak}). The lower graph shows results for a much higher energy density in lattice units $a_{\mathrm{Lat}}$. While the low-momentum scaling behavior is insensitive to these changes, only the upper graph shows the perturbative Kolmogorov scaling at high momenta. In particular, the observed high-momentum behavior is closer to the classical thermal exponent value of one, rather than to the Kolmogorov exponent for $d=4$.  

\begin{figure}
 \begin{center}
  \includegraphics[scale=0.9]{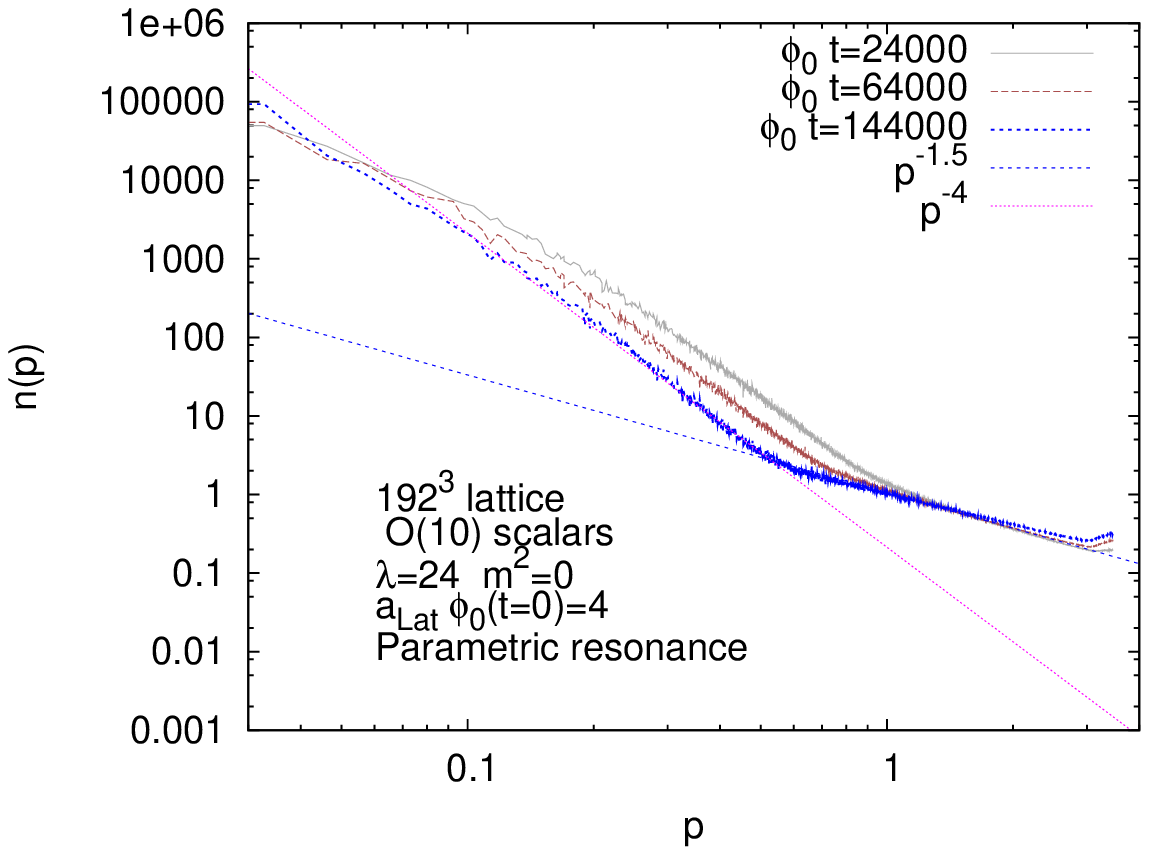}
  \caption[4d]{ The particle number spectrum of the $d=3$ simulation
for the $O(10)$-symmetric scalar field theory. }
  \label{fig:univ}
 \end{center}
\end{figure}
We have seen that, in particular, the strong turbulence regime is very insensitive to details of the underlying theory such as couplings or initial conditions. It remains to see whether there is a dependence of the scaling behavior on the number of field components $N$. In Fig.~\ref{fig:univ} we show results from simulations using $N=10$ fields in three dimensions. Again the exponents follow very closely the analytic estimates for $z=1$ and $\eta = 0$. This indicates that the universality class for the turbulent scaling exponents does not depend on $N$ in accordance with the analytic estimates.

\section{Conclusions}

Stationary turbulence is associated to conserved quantities. We have demonstrated that the nonperturbative scaling solution first observed in Ref.~\cite{Berges:2008wm} can be associated to a conserved effective particle number for the low momentum regime. The strong turbulence solution predicts a characteristic dependence on the dimensionality of space. Our classical-statistical lattice simulations provide a striking confirmation of this dependence for three and four dimensions. In particular, we see no indications for a dependence on the number of field components $N$. This strongly suggests that the universal behavior associated to the conserved effective particle number indeed only depends on the dimensionality of space and the value of the dynamic scaling exponent $z=1$ for relativistic dynamics with zero anomalous dimension.\\

We thank T.~Gasenzer for collaboration on related work. This work is supported in part by the BMBF grant 06DA9018. A large part of the numerical calculations were done
on the bwGRiD (http://www.bw-grid.de), member of the German D-Grid
initiative, funded by the {\em Bundesministerium f{\"u}r Bildung und
Forschung} and the {\em Ministerium f{\"u}r Wissenschaft, Forschung
und Kunst Baden-W{\"u}rttemberg}.

\end{document}